# LITTLE DRAGON TWO: AN EFFICIENT MULTIVARIATE PUBLIC KEY CRYPTOSYSTEM

Rajesh P Singh[1], Anupam Saikia[2] and B. K. Sarma[3]

[1]Department of Mathematics, Indian Institute of Technology, Guwahati, India
r.pratap@iitg.ernet.in
[2]Department of Mathematics, Indian Institute of Technology, Guwahati, India
a.saikia@iitg.ernet.in
[3]Department of Mathematics, Indian Institute of Technology, Guwahati, India
bks@iitg.ernet.in

## ABSTRACT

*In 1998 [8], Patarin proposed an efficient cryptosystem called Little Dragon which was a variant a variant of Matsumoto Imai cryptosystem C\*. However Patarin latter found that Little Dragon cryptosystem is not secure [8], [3]. In this paper we propose a cryptosystem Little Dragon Two which is as efficient as Little Dragon cryptosystem but secure against all the known attacks. Like Little Dragon cryptosystem the public key of Little Dragon Two is mixed type that is quadratic in plaintext and cipher text variables. So the public key size of Little Dragon Two is equal to Little Dragon Cryptosystem. Our public key algorithm is bijective and can be used for both encryption and signatures.*

## 1. INTRODUCTION

Public key cryptography has several practical applications, for example in e-commerce systems for authentication (electronic signatures) and for secure communication. The most widely used cryptosystems RSA and ECC (elliptic curve cryptosystems) are based on the problems of integer factorisation and discrete logarithm respectively. Integer factorisation and discrete logarithm problems are only believed to be hard but no proof is known for their NP-completeness or NP-hardness. Improvements in factorisation algorithms and computation power demand larger bit size in RSA key which makes RSA less efficient for practical applications. Although RSA and ECC have some drawbacks, they are still not broken. But in 1999 [1] Peter Shor discovered the polynomial time algorithm for integer factorization and computation of discrete logarithm on quantum computers. Thus once we have quantum computers in the range of 1,000 bits, the cryptosystems based on these problems can no longer be considered secure. So, there is a strong motivation to develop public key cryptosystems based on problems which are secure on both conventional and quantum computers. Multivariate cryptography is based on the problem of solving nonlinear system of equations over finite fields which is proven to be NP-complete. Quantum computers do not seem to have any advantage on solving NP-complete problems, so multivariate cryptography can be a viable option applicable to both conventional and quantum computers. MIC*, the first practical public key cryptosystem based on this problem was proposed in 1988 [5] by T. Matsumoto and H. Imai. The MIC* cryptosystem was based on the idea of hiding a monomial $x^{2^i+1}$ by two invertible affine transformations. This cryptosystem was more efficient than RSA and ECC. Unfortunately this cryptosystem was broken by Patarin in 1995[6]. In 1996 [7] Patarin gave a generalisation of MIC* cryptosystem called HFE, however in HFE the secret key computation was not as efficient as in the original MIC* Cryptosystem. The basic instance of HFE was broken in 1999[9]. The attack uses a simple fact that every homogeneous quadratic multivariate polynomial has a matrix representation. Using this representation a highly over defined system of equations can be obtained which can be solved by a new technique called relinearization [9]. Patarin [8] investigated whether it is possible to repair MIC* with the same kind of easy secret key computations. He designed some cryptosystems known as Little Dragon and Big Dragon with multivariate polynomials of total degree 2 and 3 respectively in plaintext and cipher text variables in public key with efficiency comparable to MIC*. Due to its efficiency and quadratic public key size, the Little Dragon Scheme was more interesting, however Patarin found [8], [3] that Little Dragon Scheme is insecure. Some more multivariate public key cryptosystems can be found in reference [11] and [12]. For a brief introduction of multivariate cryptography we refer to the interested readers to reference [13]. An interesting introduction of hidden monomial cryptosystems can be found in reference [3].

Designing secure and efficient multivariate public key cryptosystems continues to be a challenging area of research in recent years. In this paper we present Little Dragon Two, a modified and secure version of Little Dragon cryptosystem. Like Dragon cryptosystems the public key in our cryptosystem is of mixed type but of degree two in plaintext and cipher text variables. The efficiency of our public key cryptosystem is equivalent to that of Little Dragon cryptosystem. The complexity of encryption or signature verification is equivalent to other multivariate public key cryptosystems that is $O(n^3)$ where n is the bit size. In decryption or in signature generation we need only one exponentiation in finite field $F_{2^n}$ and this result in much faster decryption and signature generation. The outline of our paper is as follows. In section 2 we present our cryptosystem and in section 3 we give the security analysis of our cryptosystem. In section 4 we discuss its efficiency.

## 2. The Cryptosystem Little Dragon Two

Let $p$ be a prime, $n$ be a positive integer, and $F_q$ be the Galois field of $q = p^n$ elements. A polynomial $f(x)$ in $F_q[x]$ is said to be a permutation polynomial, if it is a polynomial function of $F_q$ onto $F_q$. A polynomial $f \in F_q[x]$ is a permutation polynomial of $F_q$ if and only if one of the following conditions holds:

      The function is $f$ onto;

      The function is $f$ one to-one;

$f(x) = a$ has a solution in $F_q$ for each $a \in F_q$ ;

$f(x) = a$ has a unique solution in $F_q$ for each $a \in F_q$

**Lemma 2.1[2]**

1. Every linear polynomial that is a polynomial of the form $ax + b$ with $a \neq 0$ over $F_q$ is a permutation polynomial of $F_q$.

2. The monomial $x^n$ is a permutation polynomial of $F_q$ if and only if gcd $(n, q-1) = 1$

**Lemma 2.2** The polynomial f(x) $= x^{2^r k + 2^r} + x^{2^r k} + x^{2^r}$, where $r$ and $k$ are positive integers, is a permutation polynomial of $F_{2^n}$ if and only if $2^{2^r k} + 2^r$ and $2^n - 1$ are co – prime.

**Proof.** First note that there exist integers $r$ and $k$ such that $2^{2^r k} + 2^r$ and $2^n - 1$ are co – prime. It is known that composition of two permutation polynomials is a permutation polynomial, see chapter 7 of [2]. It is easy to check that f $(x+1) = x^{2^r k + 2^r} + 1$. By lemma 2.1, $f(x+1)$ is a permutation polynomial if and only if $2^{2^r k} + 2^r$ and $2^n - 1$ are co-prime. ∎

We use $Tr(x)$ to denote the trace function from finite field, $F_{2^n} \, to \, F_2$ $i.e.$,

$$Tr(x) = x + x^2 + x^{2^2} + \ldots\ldots + x^{2^{m-1}}$$

As a consequence of above lemmas we can deduce the following lemma, which we will use to design our public key cryptosystem.

**Lemma 2.3** The polynomials $g(x) = (x^{2^r k} + x^{2^r} + \alpha)^l + x$ is permutation polynomial of $F_{2^n}$, where $Tr(\alpha) = 1$ and $l.(2^{2^r k} + 2^r) = 1 \bmod 2^n - 1$

**Proof.** First note that $Tr(x^{2^r k} + x^{2^r} + \alpha) = Tr(\alpha) = 1$, so $x^{2^r k} + x^{2^r} + \alpha \neq 0$ for all $x \in F_{2^n}$. Let $\beta$ be an element of a finite field $F_{2^n}$. Consider the equation, $g(x) = \beta$, that is,

$$(x^{2^r k} + x^{2^r} + \alpha)^l + x = \beta$$

It is clearly equivalent to

$$\left( x^{2^r k} + x^{2^r} + \alpha \right) = x + \beta$$

Raising both sides the power $2^{2^r k} + 2^r$, we get,

$$\left( x^{2^r k} + x^{2^r} + \alpha \right) = \left( x + \beta \right)^{2^{2^r k} + 2^r}$$

Or,

$$\left( x^{2^r k} + x^{2^r} + \alpha \right) + \left( x + \beta \right)^{2^{2^r k} + 2^r} = 0$$

Suppose $h(x) = \left(x^{2^{2^r k}} + x^{2^r} + \alpha\right) + (x + \beta)^{2^{2^r k} + 2^r}$. We have to show that for any $\delta \in F_{2^n}$ the equation $h(x) = 0$ has a unique solution. Note that $h(x) = 0$ and $h(x + \beta) = 0$ have the same number of solutions. Now $h(x + \beta) = 0$ is equivalent to,

$$x^{2^{2^r k} + 2^r} + x^{2^{2^r k}} + x^{2^r} + \beta^{2^{2^r k}} + \beta^{2^r} + \alpha = 0$$

Note that by lemma 2.2 $x^{2^{2^r k} + 2^r} + x^{2^{2^r k}} + x^{2^r}$ is a permutation polynomial of $F_{2^n}$. Hence the equation $h(x + \beta) = 0$ has a unique solution for any $\beta \in F_{2^n}$.  ∎

## 2.1. Public key generation

For the public key cryptosystem we cannot take all the permutation polynomials of the form $\left(x^{2^{2^r k}} + x^{2^r} + \alpha\right)^l + x$, because we want public key size to be quadratic. But the permutation polynomials in which $l$ is of the form $2^t + 1$ or $2^t - 1$ can be used to design the multivariate public key cryptosystem with quadratic public key size. For $l$ is of the form $2^t + 1$ it is not clear whether $g(x)$ is permutation polynomial or not. But for r = 0, n = 2m − 1, k = m and $l = 2^m - 1$, g(x) is permutation polynomial because in this case $2^{2^r k} + 2^r = 2^m + 1$ and $(2^m - 1)(2^m + 1) = 1 \bmod 2^n - 1$. So for public key generation we will take $g(x) = \left(x^{2^m} + x + \alpha\right)^{2^m - 1} + x$, where $\alpha$ is secret. We can take other suitable values $r, k$ and $n$ such that $l$ is of the form $2^i - 2^j$. There are few choices for $r$, $k$ and $l$ so we can assume that these are known. Suppose $s$ and $t$ are two invertible affine transformations. The relation between the plaintext and the cipher text is $g(s(x)) = t(y)$, where $x$ variable denotes the plaintext and $y$, the cipher text. Suppose $s(x) = u$ and $t(y) = v$. Thus we have the following relation between plaintext and cipher text: $\left(u^{2^m} + u + \alpha\right)^{2^m - 1} + u = v$

or $\left(u^{2^m} + u + \alpha\right)^{2^m} + (u + v) = 0$, this relation can be written as:

$$u^{2^{m+1}} + u^{2^m}v + uv + u\alpha + u^{2^m} + v\alpha + \alpha^{2^m} = 0 \qquad \textbf{(1)}$$

Suppose $B = \{\beta_1, \beta_2, \dots \dots, \beta_n\}$ is a basis of $F_{2^n}$ over $F_2$. Any $x \in F_{2^n}$ can be expressed as $x = \sum_{i=1}^{n-1} x_i \beta_i$, where $x \in F_2$. Thus $F_{2^n}$ can be identified by $F_2^n$, the set of all $n$ tuples over $F_2$. Substituting $u = s(x)$ and $v = t(y)$, where $x = (x_1, x_2, \dots \dots, x_n)$ and $y = (y_1, y_2, \dots \dots, y_n)$, we get the n quadratic polynomial equations of the form

$$\sum a_{ij} x_i x_j + \sum b_{ij} x_i y_j + \sum c_k y_k + \sum d_k x_k + e_l = 0 \qquad (2)$$

Here the coefficients $a_{ij}, b_{ij}, c_k, d_k, e_i \in \mathrm{F}_2$. Note that $u^{2^m+1}$ gives the terms of the form $\sum x_i x_j + \sum x_k + c_1$, where $c_1$ is constant and $u^{2^m}v + uv$ gives the terms of the form $\sum x_i y_j + \sum x_k + \sum y_i + c_2$, $u\alpha + u^{2^m}$ gives the terms of the form $\sum x_i + c_3$ and $v\alpha$ gives the terms of the form $\sum y_i + c_4$. So, $n$ polynomial equations of the form (2) represent the required public key.

## 2.2. Secret Key

The invertible affine transformations $(s,t)$ and finite field element $\alpha$ are secret keys.

## 2.3. Encryption

If Bob wants to send a plaintext message $x = (x_1, x_2, \ldots, x_n)$ to Alice, he does the following:

1. Bob substitutes the plaintext $(x_1, x_2, \ldots, x_n)$ in public key and gets $n$ linear equations in cipher text variables $y_1, y_2, \ldots, y_n$.
2. Second step of encryption is to solve these linear equations by Gaussian elimination method to get the cipher text $y = (y_1, y_2, \ldots, y_n)$.

## 2.4. Decryption

Here we describe the decryption algorithm.

Input: Cipher text and $y = (y_1, y_2, \ldots, y_n)$ secret parameters $(s, t, \alpha)$.

Output: Message $(x_1, x_2, \ldots, x_n)$

1: $v \leftarrow t(y_1, y_2, \ldots, y_n)$

2: $z_1 \leftarrow \alpha + 1 + v + v^{2^m}$

3: $z_2 \leftarrow z_1^{2^{m-1}}$, $z_3 \leftarrow v + 1 + z_2$

4: $X_1 \leftarrow s^{-1}(v+1)$ and $X_2 \leftarrow s^{-1}(z_3)$

5: Return $(X_1, X_2)$.

Either $X_1$ or $X_2$ is the required secret message. There are only two choices for message so it is easy to identify the correct message.

**Proof.** We prove that the procedure described above output a valid plaintext. The relation between plaintext and cipher text is $(u^{2^m} + u + \alpha)^{2^m-1} + u = v$ or equivalently $u^{2^m} + u + \alpha = (u+v)^{2^m+1}$, which can be converted into the form $(u+v+1)^{2^m+1} + v + v^{2^m} + \alpha + 1 = 0$. There are only two possibilities either $u = v+1$ or $u \neq v+1$. If $u = v+1$ then, $x = s^{-1}(v+1)$. If $u \neq v+1$ then raising both sides power $2^m - 1$ in the relation $(u + v + 1)^{2^m + 1} = v + v^{2^m} + \alpha + 1$, we get $u + v + 1 = (v + v^{2^m} + \alpha + 1)^{2^m - 1}$ or $u = v + 1 + (v + v^{2^m} + \alpha + 1)$ which implies $x = s^{-1}(v + 1 + (v + v^{2^m} + \alpha + 1))$ ∎

**Example 2.1** Here is the toy example four our cryptosystem. We are taking the finite field $F_{2^3}$ that is m = 2 and n = 3. The polynomial $x^3 + x + 1$ is irreducible over $F_2$. Suppose γ is the root of this polynomial in the extension field of $F_2$, i.e., $\gamma^3 + \gamma + 1 = 0$. Using the basis $\{1, \gamma, \gamma^2\}$ the finite field $F_{2^3}$ can be expressed as $F_{2^3} = \{0, 1, \gamma, \gamma^2, 1 + \gamma, \gamma + \gamma^2, 1 + \gamma + \gamma^2\}$. We are taking $\alpha = 1 + \gamma + \gamma^2$, as Tr $(1 + \gamma + \gamma^{2=}) \neq 0$, s(x) = $A_1 x + c_1$ and t(x) = $A_2 x + c_2$ are two invertible affine transformations, where $A_1 = \begin{pmatrix} 1 & 1 & 0 \\ 0 & 1 & 1 \\ 0 & 0 & 1 \end{pmatrix}$ and $A_2 = \begin{pmatrix} 1 & 1 & 1 \\ 0 & 1 & 1 \\ 0 & 0 & 1 \end{pmatrix}$ $c_1$ = (1, 0, 1)$^T$ and $c_2$ = (0, 1, 0)$^T$. Suppose $x \in F_{2^3}$, then x can be expressed as x = $x_1 + x_2\gamma + x_3\gamma^2$ or equivalently x = ($x_1$, $x_2$, $x_3$), where $x_i \in F_2$. Taking x = ($x_1$, $x_2$, $x_3$), we have $A_1 x + c_1$ = ($x_1 + x_2 + 1$, $x_2 + x_3$, $x_3 + 1$) and $A_2 x + c_2$ = ($x_1 + x_2 + x_3$, $x_2 + x_3$, $x_3 + x_3$). For the plaintext variable x = ($x_1$, $x_2$, $x_3$), the corresponding cipher text variable is y = ($y_1$, $y_2$, $y_3$). We have u = ($x_1 + x_2 + 1$) + ($x_2 + x_3$) γ + ($x_3$ + 1) $\gamma^2$ and v = ($y_1 + y_2 + y_3$) + ($y_2 + y_3$ + 1) γ + $y_3\gamma^2$. The relation between plaintext and cipher text is $u^{2^m+1} + u^{2^m}v + uv + u\alpha + u^{2^m} + v\alpha + \alpha^{2^m} = 0$. Substituting u and v and $\alpha = 1 + \gamma + \gamma^2$, we have the following relation between plaintext and cipher text ($x_2x_3 + x_2y_2 + x_2y_3 + x_3y_3 + x_1 + x_2 + y_1 + y_2 + y_3$) + ($x_3x_1 + x_2x_3 + x_3y_1 + x_3y_2 + x_2y_2 + x_2 + x_3 + y_2 + y_3$ + 1) γ + ($x_2x_1 + x_2y_1 + x_2y_2 + x_3y_2 + x_3y_3 + x_2 + y_3$ + 1) $\gamma^2$ = 0 or equivalently we have

$$x_2x_3 + x_2y_2 + x_2y_3 + x_3y_3 + x_1 + x_2 + y_1 + y_2 + y_3 = 0$$

$$x_3x_1 + x_2x_3 + x_3y_1 + x_3y_2 + x_2y_2 + x_2 + x_3 + y_2 + y_3 + 1 = 0$$

$$x_2x_1 + x_2y_1 + x_2y_2 + x_3y_2 + x_3y_3 + x_2 + y_3 + 1 = 0$$

Above equations represent the required public key. Note that the above equations are non- linear in plaintext variable ($x_1$, $x_2$, $x_3$) and linear in cipher text variables ($y_1$, $y_2$, $y_3$).

## 3. The Security of the proposed Cryptosystem

In this section we discuss the security of the proposed cryptosystem. In general it is very difficult to prove the security of a public key cryptosystem. For example if the public modulus of RSA is decomposed into its prime factors then the RSA is broken. However it is not proved that breaking RSA is equivalent to factoring its modulus. In this section we will give some security arguments and evidence that our cryptosystem is secure. We are using the polynomial $\left(x^{2^m} + x + \alpha\right)^{2^m - 1} + x$, where $\alpha$ is secret. Thus if we write this polynomial in the form $\sum_{i=0}^{i=d} \delta_i x^i$ then some coefficients will be 0 and 1 and some coefficients will be function of $\alpha$. Since $\alpha$ is secret so most of the coefficients of this polynomial are also secret. One important point is that the degree $d$ of this polynomial is not constant but it is function of n as m = (n + 1)/2. It is easy to see that Linearization Equation attack of [6] is not applicable to our cryptosystem. The Coppersmith–Patarin attack on Little Dragon cryptosystem [3] is due to the using monomial $x^n$ to design the little dragon cryptosystem so this attack is also not applicable to our cryptosystem. The attacks discussed in this section are Gröbner basis, univariate polynomial representation, and Differential cryptanalysis, and Relinearization, XL and FXL algorithms.

## 3.1. Attacks with Differential Cryptanalysis

Differential cryptanalysis has been successfully used earlier to attack the symmetric cryptosystem. In recent years differential cryptanalysis has emerged as a powerful tool to attack the multivariate public key cryptosystems too. In 2005 [14] Fouque, Granboulan and Stern used differential cryptanalysis to attack the multivariate cryptosystems. The key point of this attack is that in case of quadratic polynomials the differential of public key is a linear map and its kernel or its rank can be analysed to get some information on the secret key. For any multivariate quadratic function $G: F_q^n \rightarrow F_q^m$ the differential operator between any two points $x, k \in F_q^n$ can be expressed as $L_{G,k} = G(x+k) - G(x) - G(k) + G(0)$ and in fact that operator is a bilinear function. By knowing the public key of a given multivariate quadratic scheme and by knowing the information about the nonlinear part $(x^{q^l+1})$ they showed that for certain parameters it is possible to recover the kernel of $L_{G,k}$. This attack was successfully applied on MIC* cryptosystem and afterwards using the same technique Dubois, Fouque, Shamir and Sterm in 2007[16] have completely broken all versions of the SFLASH signature scheme proposed by Patarin, Courtois, and Goubin [15]. In our cryptosystem instead of using monomial of the form , $(x^{q^l+1})$ we are using the polynomial $(x^{2^m} + x + \alpha)^{2^{m-1}} + x$. Clearly the degree of this polynomial is not quadratic. Moreover the public key in our cryptosystem is of mixed type. Substituting the cipher text gives quadratic plaintext variables but in that case it will be different for different cipher texts. So to attack our cryptosystem by the methods of [14] and [16] is not feasible.

## 3.2. Univariate polynomial representation of Multivariate Public Polynomials

In our cryptosystem the encryption function is $y = t^{-1}(f(s(x)))$, where $f(x) = (x^{2^m} + x + \alpha)^{2^{m-1}} + x$. Suppose d is the degree of polynomial $f(x)$. Then $f(x)$ will give multivariate polynomials of degree w(d), where w(d) denotes the hamming weight of d. As the composition with affine transformations will not affect the degree of multivariate polynomials, so $t^{-1}(f(s(x)))$ will also give multivariate polynomials of degree w(d). Note that the degree d is not constant but it is function of n. It is easy to see that the degree of univariate polynomial representation of encryption function is not constant but it is function of n. By lemma 3.3 of [9] the degree and the number of nonzero terms of the univariate polynomial representation of encryption function are both O(n^n). The complexity of root finding algorithms e.g Berlekamp algorithm, is polynomial in the degree of the polynomial. This results in an exponential time logarithm to find the roots of univariate polynomial. Therefore this line of attack is less efficient than the exhaustive search.

### 3.3. Gröbner Basis Attacks

After substituting the cipher text in public key one can get n quadratic equations in n variables and then Gröbner basis techniques can be applied to solve the system. The classical algorithm for solving a system of multivariate systems is Buchberger's algorithm [4]. Although it can solve all the multivariate quadratic equations in theory its complexity is exponential in the number of variables. We remark that there is no closed form- formula for its complexity. In the worst case the Buchberger's algorithm is known to run in double exponential time and on average its running time seems to be single exponential (see [17]). There are some efficient variants $F_4$ and $F_5$ of Buchberger's algorithm given by Jean-Charles Faugere (see [19] and [20]). The complexity of computing a Gröbner basis by Buchberger's algorithm for the public polynomials of the basic HFE scheme is too high to be feasible. However it is completely feasible using the algorithm $F_5$. The complexities of solving the public polynomials of several instances of the HFE using the algorithm $F_5$ are provided in [10]. Moreover it has been expressed in [10] "a crucial point in the cryptanalysis of HFE is the ability to distinguish a randomly algebraic system from an algebraic system coming from HFE". Moreover our public key is of mixed type, this mean for different cipher texts we will get different system of quadratic polynomial equations, so in our public key the quadratic polynomials looks random. We are using a polynomial which has degree proportional to n. It is explained in [10] that in this case there does not seem to exist polynomial time algorithm to compute the Gröbner basis. Hence to attack our cryptosystem by Gröbner basis method is not feasible.

### 3.4. Relinearization, XL and FXL Algorithms

Relinearization, XL or FXL algorithms [9], [17] are the techniques to solve the over defined system of equations i.e., $\varepsilon$ $n^2$ equations in $n$ variables, where $\varepsilon \geq 0$. To attack the HFE cryptosystem, first the equivalent quadratic polynomial representation of HFE public key was obtained and then using the matrix representation of quadratic polynomials, they obtained $O(n^2)$ polynomial equations in $O(n)$ variables [9]. The Relinearization and XL or FXL techniques are used to solve this system of equations. Note that our polynomial is not quadratic, moreover the degree of our polynomial is not constant but it is function of n, so the attack of [9] is not feasible to our cryptosystem. Adversary cannot use directly Relinearization, XL or FXL algorithms to attack our cryptosystem because when number of equations are equal to number of variables, the complexities of these algorithms is $2^n$.

## 4. EFFICIENCY OF THE PROPOSED CRYPTOSYSTEM

In this section we give complexity of the encryption and decryption of our cryptosystem.

### 4.1. Encryption

The public key in our cryptosystem consists of n equations of the form (2). There are $O(n^2)$ terms of the form $x_i x_j$ in each n equations of the public key so the complexity of evaluating public key at message block $(x_1, x_2, \ldots, x_n)$ is $O(n^3)$. The next step of encryption is to solve the n linear equation in n cipher text variables $y_0$, $y_1$,...,$y_n$. This can be done efficiently by Gaussian elimination in $O(n^3)$ complexity. Hence the total complexity of encryption is $O(n^3)$.

## 4.2. Decryption

In the decryption of the proposed cryptosystem we need only one exponentiation namely $z_2 \leftarrow z_1^{2^{m}-1}$ . So the complexity of decryption is equivalent to Little- Dragon cryptosystem [3], [8]. Note that for exponentiation in finite fields $F_{2^n}$ there are several efficient algorithms, so the exponentiation can be performed very efficiently. The exact complexity of exponentiation will depend on the algorithm used.

## 5. CONCLUSION

We have designed an efficient multivariate public key cryptosystem. Like Little Dragon Cryptosystem the public key is mixed type but quadratic. Efficiency of our cryptosystem is equivalent to Little Dragon Cryptosystem. We have analysed our cryptosystem against all the known attacks and showed that our cryptosystem is secure.

**Authors**

Rajesh P Singh completed his MSc from Kumaun University Nainital, Uttarakhand, India in 1999. Currently, he is a PhD student in the Department of Mathematics, Indian Institute of Technology Guwahati. His research interests include Finite Fields and Cryptography.

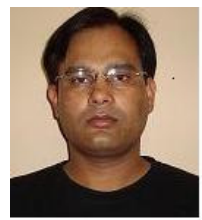

Anupam Saikia completed his PhD from University of Cambridge, UK in 2001. Currently, he is an assistant professor at the Department of Mathematics, Indian Institute of Technology Guwahati. His research interests include Elliptic Curves, Iwasawa Theory and Cryptography.

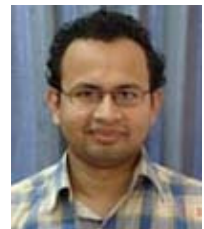

Bhaba Kumar Sarma completed his PhD in University of Delhi in 1996. Currently, he is a Professor in the Department of Mathematics, Indian Institute of Technology Guwahati. His diverse research interests include Algebraic Graph Theory, Matrix Theory and Cryptography.

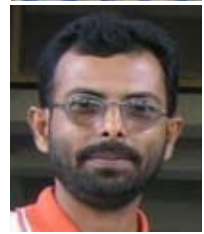